# Under Augustus' sign: the role of Astronomy in the foundation of *Augusta Praetoria Salassorum*

Stella Vittoria Bertarione
Soprintendenza per i beni e le attività culturali,
Regione Autonoma Valle d'Aosta
Ufficio Beni archeologici
Piazza Roncas, 12, 11100 Aosta, Italy.

Giulio Magli
Faculty of Civil Architecture, Politecnico di Milano
P.le Leonardo da Vinci 32, 20133 Milano, Italy.

**Abstract**

*Augusta Praetoria Salassorum* (modern Aosta) was founded around 25 BC to celebrate the victory of Augustus' army on the *Salassi*. Aosta is a "city of the founder" under many respects; for instance, one of the two twin temples of the forum was devoted to Augustus, and a huge triumphal arc to the ruler still welcomes the town's visitors. Recently, a sculpted block has been uncovered, still in its original position, on a corner of one of the town's towers. The block carries reliefs – such as a plough and a spade - which are clearly related to the town's foundation ritual. As a consequence, we carried out an archaeo-astronomical analysis of the original urban plan taking into account the complex natural horizon of the Alps in which Aosta's valley is nested. The results show that the town was very likely oriented in such a way as to pinpoint Augustus' associations with the "cosmic" signs of renewal: the winter solstice and the Capricorn.

## 1. INTRODUCTION

At the very beginning of his book on the anthropology of the urban form in Rome, Joseph Rykwert (1999) writes about the remains of Roman towns that "the more closely they are examined, the more puzzling they appear." Certainly, a reason why Roman towns can be considered "puzzling" is the fact that, although the town's foundation ritual is described by Roman historians as a rule directly inherited from the Etruscan's sacred books and closely connected with the cosmic order, only rarely have these aspects been identified in the actual archaeological records. For instance, few cases in which traces of the foundation rituals have been preserved are known (see e.g. Carandini & Cappelli 2000) and also the role of astronomical orientation is far from being assessed (Magli 2008, Gonzalez Garcia and Magli 2013). In this connection, an important aspect which awaits to be understood is the relationship – if any - between the urban plan and the cult of the founder of the town, a possibility recently identified in the urban plan of Alexandria, orientated to the rising sun on the day of birth of Alexander the Great (Bauval and Hancock 2004, Ferro and Magli 2012). In particular, we shall be concerned here with the "inventor" of the divine cult of the Roman emperor, that is, Augustus. As is well known, the symbolism connected with the new era of "Augustan peace" and the cult of the ruler's personality as keeper of the "cosmic order" are key elements of Augustus' architectural projects, such as the Mausoleum, the Meridian and the Ara Pacis (Zanker 1990, Rehak 2006). It is, therefore, natural to ask if and to what extent such ideas where reflected in the towns founded in Augustus' name and, in particular, in the foundation of the most paradigmatic of such cities, *Augusta Praetoria Salassorum* (modern Aosta). The research presented here actually started from an unexpected archaeological discovery occurred on Aosta's walls (Bertarione 2013): a sculpted corner block still *in situ.* The carvings on the block include, among others to be discussed below, a plough and a spade; it appears, therefore, to be related to the ritual of foundation of the town, so to give first hand information on such key moments. As a consequence, we have subjected the urban plan of Aosta to an accurate archaeoastronomical analysis, taking into account the (practically intact) ancient landscape in which the town was located and thus, in particular, the stunning horizon of the Alps which surrounds the town's valley.

Our results pinpoint Aosta as an example from the Roman world of a "city of the founder": indeed we show that Augustus' associations with the "cosmic" signs of renewal - the winter solstice and the Capricorn – were very probably explicitly embodied in the town's project.

## 2. THE URBAN PLAN OF *AUGUSTA PRAETORIA SALASSORUM* AND THE BALIVI TOWER

As is well known, the urban plan of the Roman colonies followed a schematic rule (see e.g. Castagnoli 1971). The city walls formed a rectangle, with the internal streets organized in an orthogonal grid. The grid was divided into four quarters by the two main roads, *Decumanus Maximus* and *Kardo Maximus.* The centre of social and religious life was preferably placed at (or near) the intersection of the main roads.

Aosta can be considered as a paradigm of this kind of urban layout (**Fig. 1**). It was founded in 25 BC shortly after the war on the *Salassi*, the original inhabitants of the region, carried out to assure a safe transit towards *Galliae* trough the Great and the Little S. Bernardo mountain pass. Especially owing to the series of studies by Rosanna Mollo (Mollo 1982a,b;1987,1988,1994,1995,1999,2000,2004,2012) we have a quite detailed understanding of the Roman town, of which much has been preserved. The walls' circuit is perfectly rectangular, with sides of 725x571 meters. Towers guarded the corners, the gates and the sides for a total of 20. One of the 4 main gates, *Porta Praetoria*, survives almost intact at the eastern entrance of the ancient town and is a masterpiece of Roman architecture; nearby lies the famous Roman theatre, with the southern façade 22 m tall. The heart of the town was in the forum, a rectangular marketplace surrounded by a covered *portico* and hosting at the centre two twin temples built on a raised terrace. The presence of the twin temples is relatively unusual; in fact in the towns founded before Augustus' rule, there was usually only one main temple, namely a *Capitolium*, that is, a replica of the main temple in Rome, where Jovis, Juno and Minerva were worshipped. Aosta's twin temples were instead almost certainly devoted to deified Rome and to Augustus, respectively.

Veneration and celebration of the founder are therefore apparent not only in the name of the town but also in his monuments, which further include a magnificent arc devoted to Augustus, built along the *Decumanus'* axis. Further, the temple's dedications appear as a clue to the fact that a more profound and intimate connection between the town and her virtual creator might have been conceived by Augustus' architects. A second, important clue to the symbolism implicit in the town's foundation has been very recently discovered during the excavations carried out at the tower located on the north-east corner of the Roman walls, known as *Torre dei Balivi* (**Fig. 2**)**.** Here an elaborately carved block has been unearthed. The block is *in situ* on the south-east corner of the tower, and was originally in plain view, the ancient street level being around one meter below (the basis of the tower was probably flooded and covered by alluvial material in the early Middle Age, thereby partly saving the sculptures from vandalism). Both sides in view are sculpted **(Figs. 3,4)**. To describe the reliefs we consider as side A the face which lies along the *Kardo* direction, and as side B that lying along the *Decumanus* direction (as verified on site, the tower is precisely oriented conformally to the walls). Side A contains two very clear reliefs, arranged into two registers: a phallus and, over it, a spade, fronting the opposite direction. Side B, also arranged into two registers, is more complex. The lower register shows again a phallus, while the upper one has a "crossed angle" figure and another, partly eroded figure to the right. The triangle-shaped object is easily and securely identified as it coincides with standard representation of the plough in the same period. Its presence on the block clearly alludes to the *sulcus primigenius* and to the town's foundation ritual. The same holds for the spade: it might be identified as the ploughshare or, with a more fascinating but equivalently feasible solution, as a Groma's peg. It is in fact extremely similar to the one which appears in the famous tomb-slab of the *gromaticus* Popidius Nicostratus from Pompeii (regarding the remaining figure, we shall come back to its possible interpretation in such a context at the end of the paper).

These elaborated reliefs are unique to the Balivi Tower and, taken as a whole, seem to show that - at the moment of the foundation - the survey operations (both symbolical and practical) were carried out starting from the point later identified with the Balivi tower's angle. Of course, the presence of sculpted symbols on Roman fortification walls is hardly a novelty. They are in fact known to occur especially in correspondence with gates or corners, however the vast majority of them are *only* phallic symbols. Such phallic symbols are ubiquitous in time and space, as they appear for instance in the walls of Republican Roman towns of central Italy like Anagni and Arpino, as well as on Hadrian's wall four centuries later. They also occur in houses' walls and public works (e.g. bridges) and almost probably had an "apotropaic" function, meaning a sort of magic defence from external agents. Much more rare are the cases in which different subjects appear in the wall's blocks, especially prepared and carefully carved before being put in place. A paradigmatic case here is in the town of central Italy called Alatri. Here carved sculptures appear in blocks set in correspondence of at least 3 gates (they are almost indecipherable today, but were documented in the 19 century) and other elaborated symbols occur in the walls of the famous Acropolis of this town. One such symbols is a composite, T-shaped figure of 3 phalli, sculpted on the lintel of the postern gate; the other, that most commentators interpret as an eagle, is on the south-east corner stone, at the basis of the most magnificent – and still intact – wall of the entire building (**Fig. 5**).

## 3. ASTRONOMY IN THE PROJECT OF *AUGUSTA PRAETORIA*

An important fact about the above mentioned Acropolis of Alatri is that, further to the carved symbols, many traces of different nature related to its symbolic foundation have been unearthed there. In particular, it is well known that a main alignment to the summer solstice sunrise was incorporated in its project (Aveni and Capone 1985, Magli 2006). Was an astronomical symbolism also connected with the project of Aosta? The answer is in the affirmative and in a sense helps us to "close the circle" about the relationship of the new town with her ideal founder.

Aosta lies at some 600 meters on sea level, nested like a jewel in the valley which bears her name. The access from the valley is from the south-west, while on all others directions prominent mountains dominate the landscape. In particular, the horizon to the east/south-east is dramatically high, being occupied by the Monte Emilius ridge which raises as much as 3000 meters at less than 9 Kms from the center of the ancient town.

The layout of Aosta roughly follows the south-west/north-east topography of the valley, but there was no special compelling constraint on the specific orientation of the axes; further, the terrain was not level and the axes do not follow the level surfaces so huge works were necessary at the moment of laying the building's blocks of the city. Since the horizon is so complex, for our data we have used a double-blind method, based on a transit survey with the horizon sampled at ½ degree intervals, validating the data with geo-referenced measures taken from Google Earth.

As far as the town's axes are concerned, the Aosta *Decumanus* bears an azimuth of 68° 02', and the *Kardo* bears the azimuth 158° 06' (transit errors are of the order of 2' or less; orthogonality of the axes thus appears to have been kept with very good precision). Taking into account the height of the horizon, it is readily seen that the sun rises along the *Decumanus* in the days around June 3 and sets along the same direction in the days around February 19 (Gregorian). In Augustus' time the calendar dates would have been the same, since the Julian calendar was perfectly tuned with the Gregorian, but the obliquity of the ecliptic has changed a bit leading to a displacement of a few days; in any case we shall not discuss further on the possible significance of these solar dates (for instance in the Roman calendar) since we are going to show that it is a pure chance that Aosta's *Decumanus* is oriented in this way. Indeed, we are going to show that its direction was fixed by the solar orientation of the *Kardo*.

At a first glance, this statement looks as a contradiction in terms, and indeed, the amplitude in degrees spanned by the rising sun with a flat horizon at the latitudes of northern Italy is about 35° north/south of east. This means that, usually, if a direction is oriented to the rising sun, the direction orthogonal to it is not. This situation may, however, change drastically if the horizon is very high, since the horizon can "delay" the effective appearance of the sun until it reaches a sensible height and, consequently, a higher azimuth. This is precisely what happens in Aosta at the winter solstice: the sun rises with a theoretical azimuth of 125° but remains behind the mountain ridge to the south-east until it has an altitude of 17°. The azimuth of the sun at this altitude is very close to that of the Aosta *Kardo* (**Fig.6**) (to be precise, today the winter solstice sun has an altitude of 17° 20' at this azimuth, while in Augustus' times – due to the slight variation of the obliquity of the ecliptic – the altitude was about 25' greater). Putting it in other words, Aosta is oriented in such a way that its axis closer to the meridian points to the rising sun at the winter solstice; as a necessary consequence, also the complementary axis is aligned with the sun.

## 4. DISCUSSION

We strongly support the idea that orientation to the winter solstice was deliberate. The winter solstice was indeed hosted by the sign of Capricorn at those times (actually, precession was moving it in Sagittarius; today it is in the region of Ophiucus, between Sagittarius and Scorpio). As is well known, astrology was a key instrument of Augustus *propaganda* and Capricorn was a "logo", an emblem chosen to legitimate and iconize his new order (Barton 1995). Many debates exist in the literature as to why he choose this sign – which was not his birth sign, as he was born on the 23 September, thus close to the autumn equinox in Libra. However, the choice of Capricorn, and more specifically the winter solstice, was easily referred to by Augustus in astrological term as being the sign of his conception. Capricorn clearly fitted much better than Libra with the idea of renewal, traditionally associated with the midwinter sun. It was thus chosen to signify the new "golden era" of peace and prosperity, as testified in uncountably many iconographic sources (such e.g. coins and cameos like the *Gemma Augustea*). Therefore an orientation of "newborn" Augusta to the sun rising on the winter solstice in the sign of her founder is pretty conceivable. In this connection, we would like to propose – at the level of educate guess - that the image difficult to read located on side B of the cornerstone might have been a Capricorn as well, as hinted at also by a comparison with known representations of Augustan times (**Fig. 7**). To conclude, the astronomical orientation of *Augusta Praetoria* was likely conceived to put in evidence the relationship of the town with her founder Augustus, fixing a fascinating appointment between the midwinter sun and the rigorous geometry of the urban layout, framed within the scenario of the Italian Alps.

**ACNKOWLEDGMENTS**
The research presented here has been made possible by the continual support of the *Soprintendenza per i beni e le attività culturali* of Regione Autonoma Valle d'Aosta. In particular the authors gratefully acknowledge the superintendent Architect Roberto Domaine and the director for restoration and promotion, Architect Gaetano De Gattis. G.M. further acknowledges support by FARB (University fund for basic research) at the Politecnico of Milan.

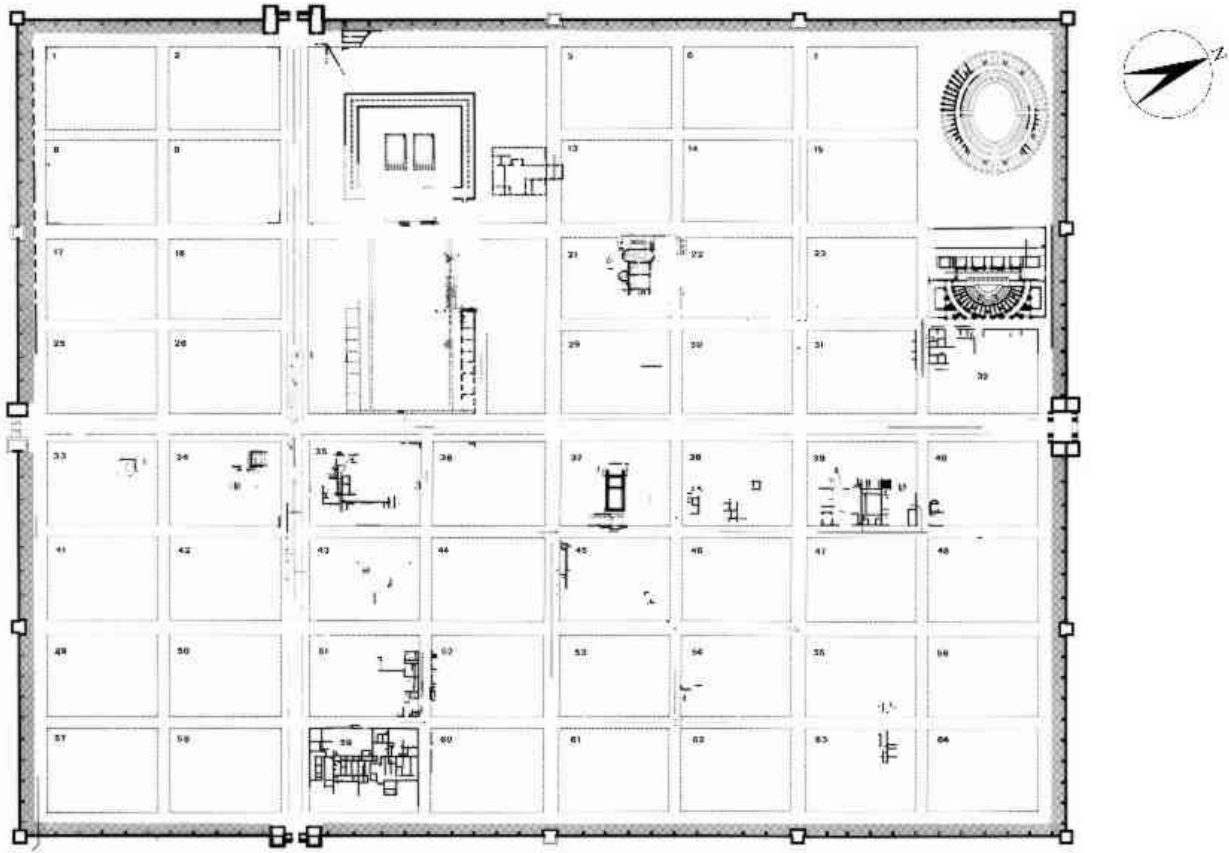

**Fig. 1   Plan of *Augusta Praetoria Salassorum*, Roman Aosta.**

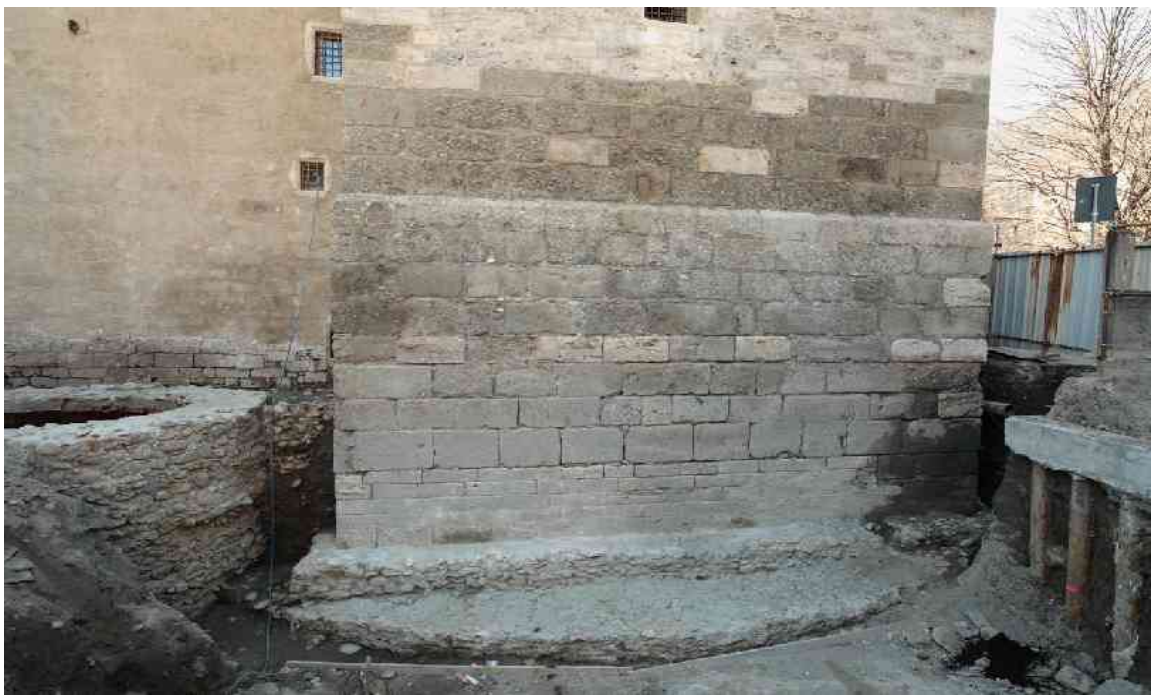

**Fig. 2 Aosta. The Balivi tower at the end the recent excavations. View from the east (Photograph S. Bertarione)**

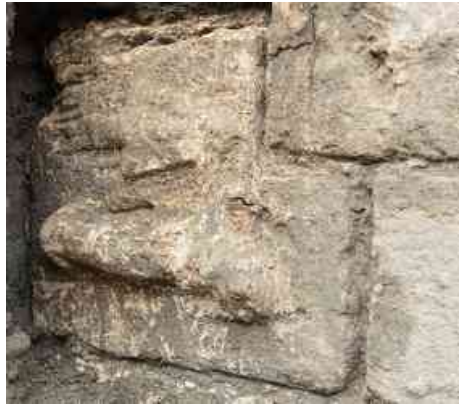

**Fig. 3 Aosta. The corner block relief discovered in the Balivi Tower, side A (Photograph S. Bertarione)**

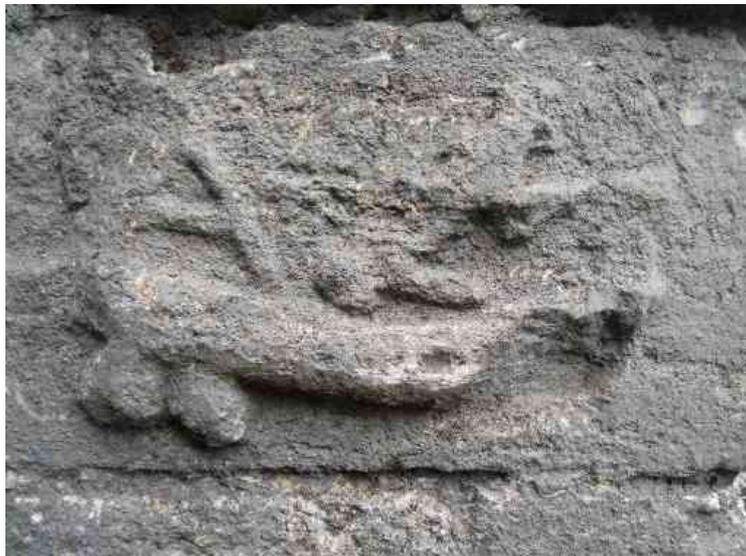

**Fig. 4 Aosta. The corner block relief discovered in the Balivi Tower, side B (Photograph S. Bertarione)**

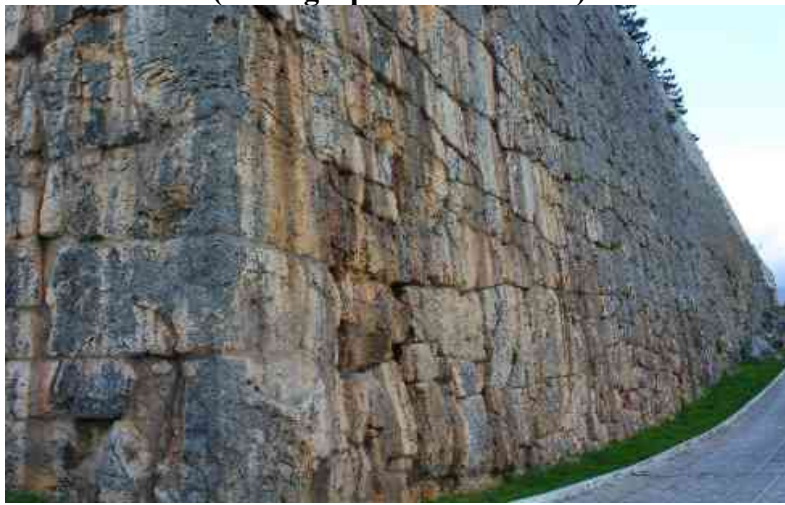

**Fig. 5. Alatri. The south-east corner of the Acropolis with the corner block relief at the basis. (Photograph G. Magli)**

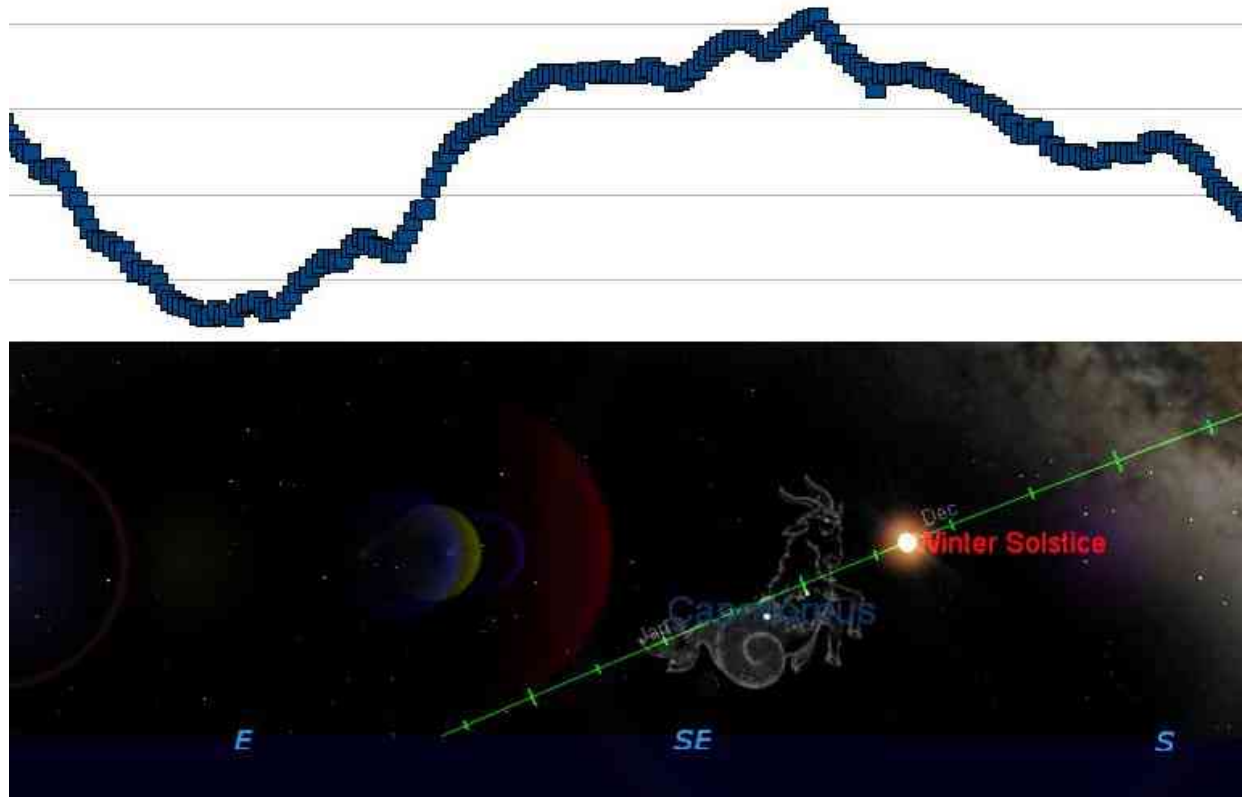

**Fig. 6 Aosta. A computer-simulated path of the rising sun at the winter solstice in 25 BC compared with the corresponding profile of the horizon (up).**

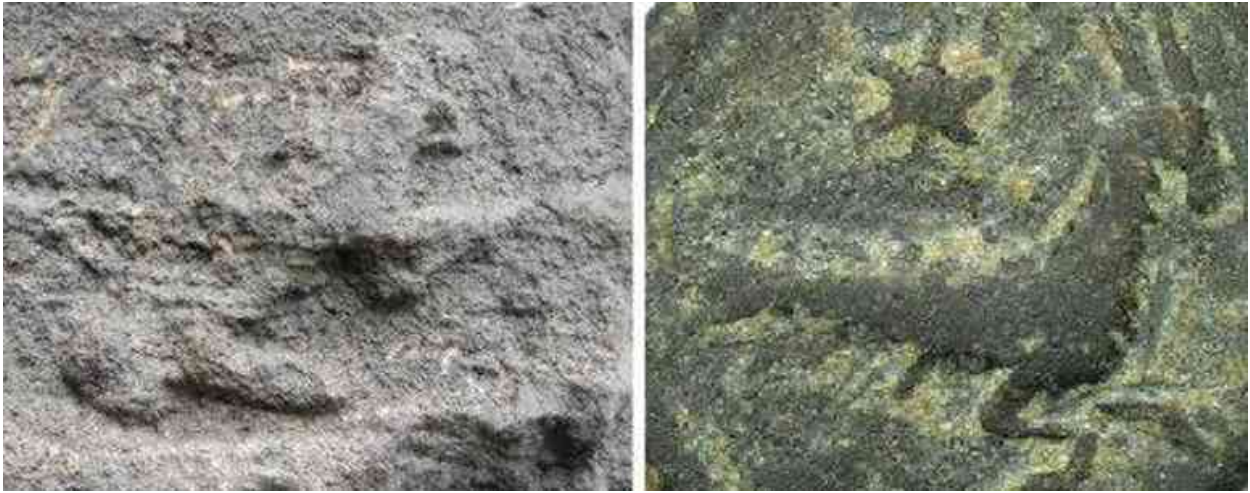

**Fig. 7 The image of Capricorn on a coin of Augustan period (right) and a close-up of the relief on side B of the corner block on the Balivi tower (left).**